# INTERPRETABILITY IS NOT EXPLAINABILITY: NEW QUANTITATIVE XAI APPROACH WITH A FOCUS ON RECOMMENDER SYSTEMS IN EDUCATION

Riccardo Porcedda

Universita degli Studi di Milano-Bicocca, Milan, Italy

## ABSTRACT

*The field of eXplainable Artificial Intelligence faces challenges due to the absence of a widely accepted taxonomy that facilitates the quantitative evaluation of explainability in Machine Learning algorithms.*

*In this paper, we propose a novel taxonomy that addresses the current gap in the literature by providing a clear and unambiguous understanding of the key concepts and relationships in XAI. Our approach is rooted in a systematic analysis of existing definitions and frameworks, with a focus on transparency, interpretability, completeness, complexity and understandability as essential dimensions of explainability. This comprehensive taxonomy aims to establish a shared vocabulary for future research.*

*To demonstrate the utility of our proposed taxonomy, we examine a case study of a Recommender System designed to curate and recommend the most suitable online resources from MERLOT. By employing the SHAP package, we quantify and enhance the explainability of the RS within the context of our newly developed taxonomy.*

## 1. INTRODUCTION

The growing prominence of Artificial Intelligence in various domains, including education, has led to an increased focus on building reliable and human-centered applications. In parallel, the challenge of explainability in Machine Learning algorithms remains an open problem. Despite a burgeoning body of research in the XAI field, proposed solutions often lack clarity, sufficiency, or efficiency, with results frequently based on user tests related to multiple and often correlated concepts such as trust, transferability, and informativeness [1]. Moreover, the current scientific literature does not provide a consensus on the distinction between explainable and interpretable systems, with only a limited number of papers addressing this issue.

Gilpin, Leilani H. et al. [2] attempt to address this problem by focusing on the concepts of interpretability and completeness as essential components of explainability:

- Interpretability refers to a system's ability to describe its inner workings in a manner intelligible to humans;
- Completeness entails accurately describing a system's operations.

For instance, an explanation is deemed complete if users can understand it well enough to predict the behavior of the system described. A naive solution might involve presenting the entire set of operations performed by the system; however, users could become overwhelmed by the complexity of the inner mechanism, losing interpretability.

Consequently, Gilpin et al. [2] view explainability as a trade-off between interpretability and completeness.

Nevertheless, Herman [3] cautions that adopting such an anthropocentric taxonomy may be hazardous, as human evaluation can harbor biases that turn any persuasive system into an explainable one, regardless of the truth-value of the explanations provided.

Despite these challenges, explainability necessitates a human-in-the-loop process, as the human user is the ultimate recipient of explanations. In this paper, we propose a taxonomic classification of explainability-related concepts that minimizes, yet preserves, the essential involvement of the human user. This approach enables the construction of a pipeline for implementing this framework in AI applications, such as intelligent information systems and soft computing techniques. We then demonstrate the practical application of this pipeline in a case study of a Recommender System, which takes resources and relative ratings datasets from MERLOT, user metadata preferences, and produces binary classification outputs of the items (recommended/not recommended). This example underscores the relevance and potential impact of our proposed taxonomy on the development and understanding of AI systems across various domains.

## 2. RESEARCH METHODOLOGY

### 2.1. A New Taxonomy Proposal

We now propose a new XAI taxonomy to tackle the issue presented above, and we begin by unraveling the notions of interpretability and explainability into more specific definitions of transparency, interpretability, explanation, completeness, complexity, understandability, and explainability.

This is necessary to provide a formal mathematical framework and use it to create quantitative measures.

Transparency is the cornerstone of explainability: it tells us how much opaque is a model and, from that, it is possible to decide how to explain it. Lipton [4] highlights three key aspects of transparency: *simulatability*, *decomposability*, and *algorithmic transparency*. Informally, transparency is the opposite of opacity or *black-box-ness*, indicating an understanding of the mechanism by which the model works.

Simulatability In the strictest sense, a model can be considered transparent if a person can contemplate the entire model at once. This definition suggests that an interpretable model is a simple model, allowing humans to grasp its decision-making process more easily.

Decomposability A second notion of transparency is decomposability, which means that each part of the model - each input, parameter, and calculation - admits an intuitive explanation. This aspect of transparency allows users to understand the model's components and how they contribute to the overall decision-making process.

Algorithmic Transparency A final notion of transparency might apply at the level of the learning algorithm itself. Algorithmic transparency refers to the understandability of the algorithm used to train the model, allowing users to comprehend the rationale behind the model's development and behavior.

In our taxonomy, we define transparency in the following way:

Definition 1 (Transparency). *A system is transparent if its inner mechanism is accessible. By inner mechanism, we mean:*

−       *The definition of the algorithm that makes the system work;*
−       *The reconstruction, given an output of the system, of all operations executed by the algorithm itself.*

*A system with maximum transparency is called a white-box. A system with zero transparency is called a black-box.*

This definition encompasses the aspects of simulatability, decomposability, and algorithmic transparency discussed by Lipton [4].

Before the work of Gilpin et al. [2], Doshi-Velez, F. and Kim, B. [5] emphasize the need for a rigorous science of interpretability in machine learning and propose a taxonomy of interpretability methods. They defined interpretability as the ability to explain or to present in understandable terms to a human.

But, as the authors themselves stated, a formal definition of explanation remained elusive. Furthermore, they run in a cyclic definition where interpretability, an ingredient of explainability, needs the definition of explanation to work.

In this paper, we follow the pathway shown by Gilpin et al. [2] to get the following definition: Definition 2 (Interpretability). *A system is interpretable if it is possible to identify a statistical or functional relation $R$ between inputs and outputs.*

*The relation can be represented as a mathematical function, a transformation, a rule, or a set of rules in the IF-THEN form.*

*By definition, a white-box system has maximum interpretability, and a black-box has zero interpretability.*
It is possible for black-box systems to be approximated by white-box systems. In that case, the interpretability $I(R)$, which can be thought of as a real value in the interval $[0,1]$, should depend on these two features:

−       The approximation method (global or local);
−       The fidelity measure, which is how well the white-box system approximates the black-box.

The definition of explanation given by the Cambridge Dictionary of English Language is "*the details or reasons that someone gives to make something clear or easy to understand*". Adapting this definition to our purpose

**Definition 3 (Explanation).** *An explanation X is a reason or justification given for one or more outputs of a system.*

*The explanation can be given in any preferred form (textual, graphical, etc.), but it must be intelligible.*

**Definition 4 (Completeness).** *The completeness C(X) of an explanation X measures its adherence to the interpretation R .*

For instance, if a given classifier provides the classification rules as explanations, we have maximum completeness. Miller [6] highlights the importance of completeness in explanations, stating that a complete explanation should provide all the necessary information to understand a model's decision-making process.

To define understandability later, we introduce the concept of Complexity.

**Definition 5 (Complexity).** *The complexity ω(X) of an explanation X measures the effort required to understand it.*

For instance, considering IF-THEN classification rules as explanations, we propose a measure for their complexity. It should satisfy these requirements:

1. The effort to understand a classification rule increases with the number of IF-THEN clauses;
2. Rules with more than one IF-THEN clause on the same feature are more complex;
3. A rule with only one IF-THEN clause has the lowest complexity (0).

We propose the following formula for the complexity of the *i*-th rule $R_i$:

$$\omega_i = \alpha_i(||R_i|| - 1) = \frac{||R_i||}{\#S_i}(||R_i|| - 1), \quad S_i \subseteq K \tag{1}$$

where:

  − $||R_i||$ is the length of the *i*-th rule $R_i$ (number of IF-THEN clauses concatenated with an AND); − $K$ is the set of features used in the model;
  − $S_i$ is the subset of features on which IF-THEN clauses of the rule $R_i$ are executed.

This linear function of the number of IF-THEN clauses satisfies the requirements. For a set of rules, the total complexity could be the sum or the average of individual complexities.

R. Guidotti et al. [7] repeatedly emphasize how important it is to have understandable explanation in order to obtain interpretable algorithms. To address the definition of Understandability, we refer to the work of Arrieta, A. B. et al. [8]: it denotes the characteristic of a model to make a human understand its function − how the model works − without any need for explaining its internal structure or the algorithmic means by which the model processes data internally.
We synthesize and insert this definition in our framework by stating the following:

Definition 6 (Understandability). *Understandability U of an explanation X represents the user's capacity to reproduce the system's behavior as described by the explanation itself.*

This definition introduces subjectivity, as different users might have varying levels of understanding.

The understandability depends on two factors:

–        Complexity of the explanation $\omega(X)$
–        Tolerable complexity $\omega_b$ for the user

We desire a function that meets the following conditions:
–        Has domain $[0, +\infty)$ and range $(0, 1]$
–        $U(\omega = \omega_b; \omega_b) \simeq 0.9$
–        $U(0; \omega_b) = 1$
–        $\lim_{\omega \to \infty} U(\omega; \omega_b) = 0$

Additionally, we desire a "sigmoidal" behaviour, with:

–    $\frac{dU(\omega)}{d\omega} \simeq 0$ for both small (range of *learning ease*) and great (range of *saturation*) values of $\frac{\omega}{\omega_b}$;
–    $-1 \leq \frac{dU(\omega)}{d\omega} < 0$ for intermediate values of $\frac{\omega}{\omega_b}$ (range of learning decline).

We propose two candidate functions.

**Gaussian decline**

$$U(\omega; \omega_b) = exp\left(-\left(\frac{\omega}{3\omega_b}\right)^2\right) \qquad (2)$$

**SHT (Squared Hyperbolic Tangent) decline**

$$U(\omega; \omega_b) = 1 - tanh^2\left(\frac{\omega}{3\omega_b}\right) \qquad (3)$$

Both functions satisfy the desired properties listed above, and their difference lies in how aggressive the decline is (how negative the derivative is for intermediate values of $\frac{\omega}{\omega_b}$)

Lipton [4] discusses the trade-off between complexity and understandability in machine learning models, arguing that the ideal explanation should balance these aspects. By incorporating these two factors, our definitions of completeness, complexity, and understandability aim to provide a more comprehensive framework for evaluating the quality of explanations in the context of machine learning.

Additionally, Arrieta, A. B. et al. [8] state that explainability is associated with the notion of explanation as an interface between humans and a decision maker that is, at the same time, both an accurate proxy of the decision maker and comprehensible to humans. Putting everything together, we get to the following definition:

Definition 7 (Explainability).

*A system is explainable if it is capable of providing explanations of its inner mechanism which are both complete and understandable.*

Operationally, the Explainability $E(R, X)$ of a system is here intended as the product of its Interpretability $I(R)$, the Completeness $C(X)$ and the Understandability $U(X)$ of the explanations:

$$E(\mathcal{R}, X) = I(\mathcal{R})C(X)U(X) \tag{4}$$

## 2.2. How to improve the Explainability of a Machine Learning model

If an explanation is not considered sufficient by the user, there is only one way to improve it: providing additional explanations.

We should have a measure of the Total Explainability for a given number of explanations $X_i$ with Explainability $E_i$.

Total Explainability of Two Explanations First, we begin by defining the desired properties of the Total Explainability $Tot(E_1, E_2)$ of two explanations:

1. $Tot(E_1, E_2) = Tot(E_2, E_1)$ (Symmetry)
2. $\max(E_1, E_2) \leq Tot(E_1, E_2) \leq 1 \qquad \forall E_1, E_2 \in [0, 1]$
3. $Tot(E_1, 0) = E_1 \qquad \forall E_1 \in [0, 1]$
4. Let be $E_1, E_2, E_3 \in [0, 1]$ such that $E_1 \leq E_2 \leq E_3$, then $Tot(E_1, E_2) \leq Tot(E_2, E_3)$ (Monotony)

For the symmetry axiom,

$Tot(E_1, E_2) = Tot(E_2, E_3)$ if and only if $E_1 = E_3$.

We propose the candidate solution:

$$Tot(E_1, E_2) = \max(E_1, E_2) + [1 - \max(E_1, E_2)]\min(E_1, E_2) \tag{5}$$

It is trivial to demonstrate that properties 1, 2, and 3 are satisfied. Here, we attach the proof for property 4.

Theorem 1. *Equation 5 satisfies Property 4.*

*Proof.* Let be $E_1, E_2, E_3 \in [0, 1]$ such that $E_1 \leq E_2 \leq E_3$ and we define a variable

$S = Tot(E_2, E_3) - Tot(E_1, E_2)$. If Property 4 is satisfied, $S \geq 0$.

$S = \max(E_2, E_3) + [1 - \max(E_2, E_3)]\min(E_2, E_3) - \max(E_1, E_2) + [1 - \max(E_1, E_2)]\min(E_1, E_2) =$

$$= E_3 + (1 - E_3)E_2 - E_2 - (1 - E_2)E_1 =$$
$$= E3 - E2E3 - E1 + E1E2 =$$
$$= (E_1 - E_3)(E_2 - 1)$$

But $E_2 \in [0, 1]$ and $E_1 \leq E3$, hence

$$S = (E_1 - E_3)(E_2 - 1) \geq 0$$

Recursive Definition of Total Explainability for $k$ Explanations Once we defined the Total Explainability of two explanations, we can write a recursive definition for a set of $k$ explanations: the idea behind this is that, for any $k$, the Total Explainability can be computed on $E_k$ and the set of $E_{\{k-1\}}$ explanations.

$$E_{\{k\}} = \begin{cases} E_1 & if \quad k = 1 \\ Tot(E_k, E_{\{k-1\}}) & if \quad k \neq 1 \end{cases} \qquad (6)$$

## 2.3. Choice of Explainability-Augmentation Method and Related Metrics

After evaluating the *pros* and *cons* of the most promising methods found in the literature, such as SHAP, LIME, and counterfactuals, we have chosen to adopt the SHAP approach. This decision is based on its ease of implementation, computational efficiency, and the provision of intuitive graphical explanations.

Shapley values, a concept borrowed from Game Theory, are designed to represent rewards assigned to players in a coalition based on their marginal contribution to the total coalition payoff [9]. In this context, the game represents our machine learning model's classification task, the players are the features used, and the total coalition payoff is the classification value. The Shapley value of a feature is defined as follows:

$$\phi_j(val) = \sum_{S \subseteq 1,...,p \setminus j} \frac{|S|!(p - |S| - 1)!}{p!}(val(S \cup j) - val(S)) \qquad (7)$$

Here, *val* is a characteristic function that expresses the utility for any subset $S$ of the $p$ features of the model. For a given instance $x$, the value is calculated as:
Z

$$val_x(S) = \int \hat{f}(x_1, ..., x_p)d\mathbb{P}_{x \notin S} - E_X(\hat{f}(X)) \qquad (8)$$

This represents the prediction for feature values in $S$, marginalized over features not in $S$.

The innovation of SHAP (SHapley Additive exPlanations) [10] lies in its use of Shapley values to build an explanation model $g$ that is additive with respect to the features of the original model:

$$g(z') = \phi_0 + \sum_{j=1}^{M} \phi_j z'_j \qquad (9)$$

The steps of the SHAP method are as follows:

1. Sample coalition vectors $z'$ (binary vectors indicating the presence/absence of a feature)

2. Convert vectors $z'$ to the original feature space and obtain predictions from the original model; for features with a zero-value in the coalition vector, the conversion to the feature space is performed by substituting a random value from the data

3. Compute the weight of each coalition vector using the *SHAP kernel*

4.  Fit model *g* and return the coefficient values (which are, in fact, the Shapley values)

To conduct an offline evaluation, we require a quantitative explainability measure based on Shapley values. A list of *mask-based* and *resample-based* metrics is proposed by Lundberg et al [11].

Mask-based metrics *Mask-based* metrics allow observation of the model's output changes when features are masked with their mean value. For instance, the *Keep-positive* (mask) metric retains features with the most positive Shapley values, masking the others (features with negative Shapley values are always masked) for each instance and for an increasing fraction of features. Plotting the fraction of features kept versus the model output produces a curve that measures how well the local explanation method has identified features that increase the model's output for this prediction. Higher valued curves represent better explanations (an example is shown here in Figure 1).

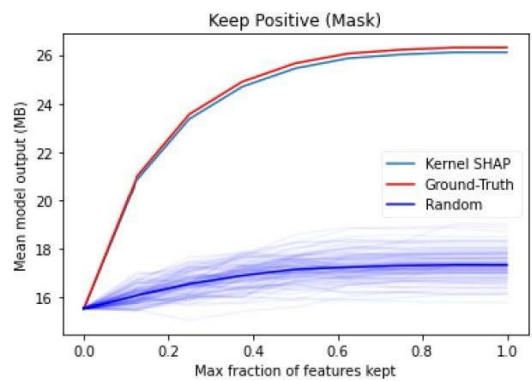

Figure 1: Illustrative example of the *Keep-positive* metric [12].

Resample-based metrics The Keep Positive (resample) metric is similar to the Keep Positive (mask) metric, but instead of replacing hidden features with their mean value, it replaces them with values from a random training sample. If the input features are independent, this estimates the expectation of the model output conditioned on the observed features. The mask-based metric can also be viewed as approximating the conditional expectation of the model's output, but only if the model is linear. The resample-based metric does not make the assumption of model linearity.

Keep-absolute (resample) In our classification task, we are interested in both positive and negative *Shapley-valued* features, so we need a single metric that captures the quality of the explanations provided considering the overall feature importance.

The Keep Absolute metric measures the explanation method's ability to find the features most important for the model's accuracy. It operates similarly to the Keep Positive metric, but keeps the most important features according to the absolute value of the associated Shapley values. Since removing features based on their absolute effect on the model does not specifically push the model's output higher or lower, we measure not the change in the model's output, but rather the change in the model's accuracy. Good explanations will enable the model to achieve high accuracy with only a few important features.

As a proxy for Explainability, we can rely on the Area Under Curve (AUC) of the curve plotted. This allows us to quantitatively assess the quality of our explanations and make improvements where necessary.

# 3. IMPLEMENTATION EXAMPLE ON IRIS DATASET WITH SVM MODEL

To easily show the results of our approach, we first implement it on a toy dataset and, after that, we show the result on our RS.

Using the *sklearn* Python library, we train a Support Vector Machine (SVM) Classifier on the Iris Dataset.

We then extract classification rules with the SVM+Prototype algorithm [13].

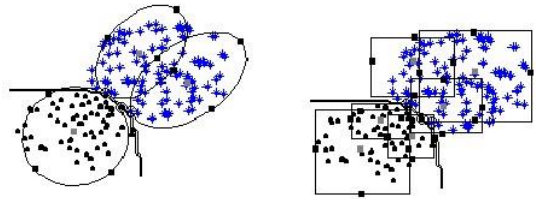

Figure 2: Ellipsoids for equation rules and Hyper-rectangles for interval rules (Illustrative example of the SVM+Prototype Algorithm).

The algorithm works as follows:

1. Train an SVM classifier on the dataset.

2. Identify the support vectors, which are the critical data points that define the decision boundarybetween classes.

3. For each support vector, find its closest prototype (representative) from the same class. Aprototype can be a centroid or medoid of the class, or another representative point.

4. Define hyper-rectangles or ellipsoids around each support vector and its associated prototype.These geometric shapes are used to form the rules. The decision boundary of the SVM is approximated by the union of these shapes.

5. Extract the rules from the geometric shapes (see Figure 2). For hyper-rectangles, the rulesare formed as conjunctions of intervals for each feature, while for ellipsoids, the rules are represented as quadratic equations.

Let us consider an example of rule extraction on the Iris dataset using the SVM+Prototype algorithm. The Iris dataset has four features: Sepal Length, Sepal Width, Petal Length, and Petal Width. We will focus on extracting rules for distinguishing between the Iris Setosa and Iris Versicolor classes.

Assume that we have trained an SVM on the Iris dataset and identified the support vectors. Suppose we find the following support vector and its closest prototype for Iris Setosa:

–     Support Vector: (5.1, 3.5, 1.4, 0.2)
–     Prototype: (5.0, 3.4, 1.5, 0.3)

Next, we define a hyper-rectangle around the support vector and its prototype:

−    Sepal Length: [5.0, 5.1]
−    Sepal Width: [3.4, 3.5]
−    Petal Length: [1.4, 1.5]
−    Petal Width: [0.2, 0.3]

The extracted rule from this hyper-rectangle is: *IF 5.0 ≤ Sepal Length ≤ 5.1 AND 3.4 ≤ Sepal Width ≤ 3.5 AND 1.4 ≤ Petal Length ≤ 1.5 AND 0.2 ≤ Petal Width ≤ 0.3 THEN Iris Setosa*

The same process is applied to other support vectors and their prototypes, generating multiple rules for each class.

The average complexity of the generated rules are calculated with Eq.1 and its value is $\omega = 3$.

To get the Explainability of this model, we should first notice that a SVM is a white-box model (therefore, maximally interpretable) and, if we provide the generated rules as explanations, it yields total completeness. All of this means that Eq.4 is simplified as

$$E(X) = U(X) \qquad (10)$$

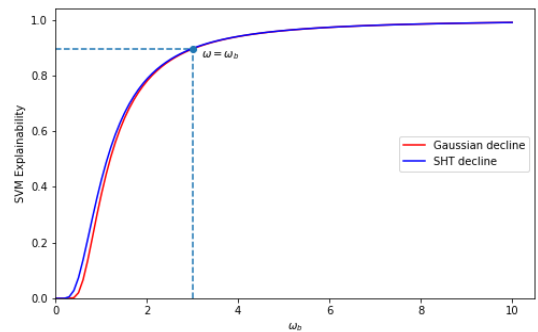

Figure 3: Model Explainability as a function of $\omega_b$.

Because in this paper we do not estimate the parameter $\omega_b$, we show how the SVM Explainability changes as a function of the same parameter.

It is possible to see (as desired when we defined the properties of the Understandability measure), that $E(X) \simeq 0.9$ when $\omega = \omega_b$.

Now we improve this result adding explanations with SHAP (Kernel Explainer).

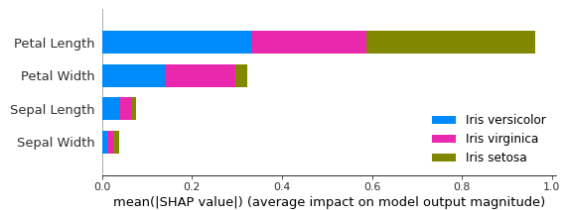

Figure 4: Importance plot of Iris features.

In Figure 4 the user can already get some valuable insights:

−       Petal Length is the most important feature;
−       Sepal Width is the least important feature;
−       The contributions of Sepal Width, Sepal Length and Petal Width to the classification of the Iris setosa are close to zero.

Proceeding with the quantitative evaluation of the Explainability with the new results from SHAP, we compute and show the Keep Absolute (resample) metric.

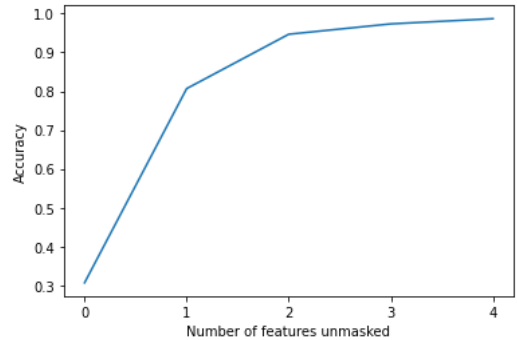

Figure 5: Keep absolute (resample) metric on SVM model for Iris dataset.

The quantitative evaluation confirms what was already shown in the Importance Plot: even unmasking only the two most important features (Petal Width and Petal Length) we get an accuracy of 95%.

The AUC of the above curve is equal to 0.69.
Now it is possible to compute and show the Total Explainability of the SVM using the methods illustrated in section 3.

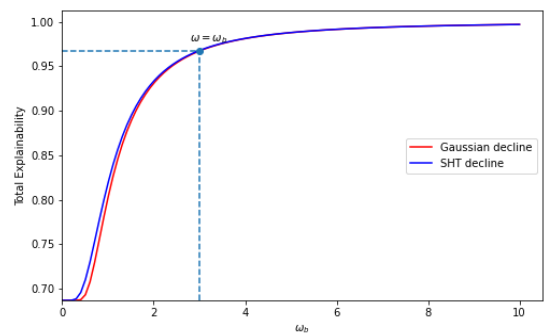

Figure 6: Total Explainability of the SVM model as a function of $\omega_b$.

The difference between Figure 5 and Figure 6 is that the curve in the latter plot starts from a total Explainability equal to the AUC of the Keep absolute metric curve: the Explainability has been greatly increased, even for low values of the bearable complexity $\omega_b$.

## 4. IMPLEMENTATION ON MERLOT DATASET WITH RECOMMENDER SYSTEM

After a brief illustration of the potential of our augmentation methods on a toy dataset, we proceed with the evaluation and augmentation of the Explainability of the Recommender System with a detailed description of every step.

### 4.1. Data

Before discussing the implementation details, let us briefly introduce the MERLOT dataset.
The MERLOT dataset is a collection of educational resources and user ratings, widely used in the context of Recommender Systems for online learning. It contains metadata about various courses and user-generated reviews and ratings, making it a valuable source of information for building personalized recommendations in the educational domain.

We work with two MERLOT datasets: resources (Table 2) and ratings (Table 1).

The first dataframe contains all metadata regarding the courses, while the second contains all reviews and ratings submitted by users (in both tables, only useful features are shown). The *resource* column in this last dataframe corresponds to the *id* column in the first one.

Then, to create the starting point of the dataset on which our RS is trained, we merge the datasets by right-join on *resource = id*.

We then simulate a user with the following preferences:

| | |
|---|---|
| disciplines | Business |
| language | English |
| difficulty | Medio Alta, Media |
| duration | 0-30, 30-60 |
| format | Text, Video |
| type | Simulation, Tutorial |
| min_age | 0 |
| max_age | 100 |

Table 1: First rows of MERLOT ratings dataframe

| Id | rating | resource |
|---|---|---|
| 0 | 4 | 0 |
| 1 | 3 | 0 |
| 2 | 5 | 0 |
| 3 | 3 | 1 |
| 4 | 5 | 1 |

Table 2: First rows of MERLOT resources dataframe.

| id | type | language | difficulty | format | duration | min_age | max_age | discipline_level_0 | discipline_level_1 | discipline_level_2 | discipline_level_3 |
|---|---|---|---|---|---|---|---|---|---|---|---|
| 0 | Presentation | English | Medio Bassa | Text | 0-30 | 18 | 20 | Business | Economics | Micro | Absent |
| 1 | Simulation | English | Medio Alta | Website | 0-30 | 14 | 20 | Humanities | History | Area Studies | Africa |
| 2 | Tutorial | English | Bassa | Text | 30-60 | 14 | 20 | Mathematics and Statistics | Mathematics | Geometry and Topology | Euclidean Geometry |
| 3 | Simulation | English | Medio Bassa | Text | 0-30 | 18 | 20 | Science and Technology | Physics | Electricity and Magnetism | Circuits |
| 4 | Simulation | English | Medio Bassa | Text | 0-30 | 18 | 28 | Science and Technology | Physics | Mathematics | Absent |

## 4.2. Feature Correlation

Before proceeding to the explanation through KernelSHAP, we must check for feature correlation: this could lead to unreliable computations of Shapley values.

Since we are working with both numerical and text data, we compute the correlation matrix using Pearson correlation for numerical features and Cramer's V for categorical features.

Pearson correlation measures the linear relationship between two continuous variables, while Cramer's V measures the strength of association between two categorical variables.

In Figure 7 we can notice a correlation cluster corresponding to the discipline levels, with significantly high value ($V = 0.96$ for discipline levels 0 and 1).

To mitigate this problem, we aggregate disciplines levels and, to get the Shapley values, we will only consider the resultant feature *disciplines*.

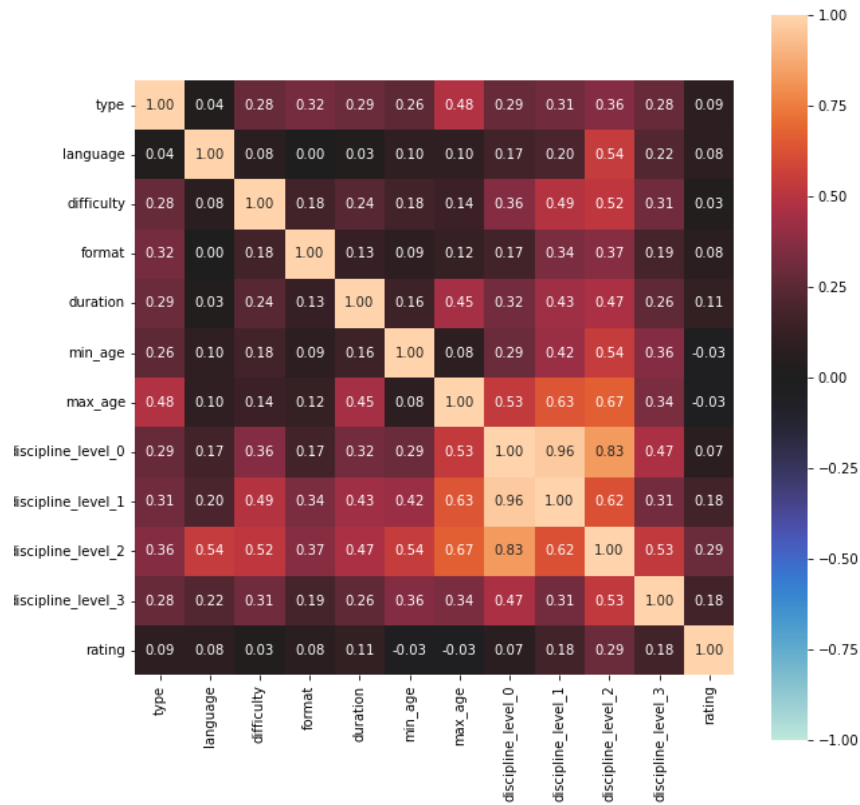

Figure 7: Correlation matrix of training set.

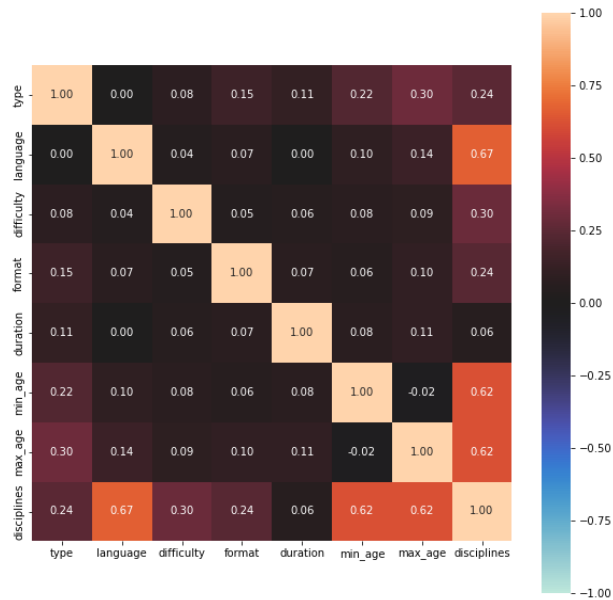

Figure 8: Correlation matrix of the background dataset.

There are still a few notable correlations, but the highest value does not go beyond 0.67.

The idea behind grouping and ungrouping discipline levels is to manage the feature correlation by aggregating them into a single feature. By doing so, we can compute Shapley values for the aggregated feature *disciplines* (which are called Owen values). This approach helps in mitigating the high correlation between discipline levels, which could lead to unreliable Shapley value computations.

## 4.3. Predict Function and Encoding

Kernel Explainer needs basically two things: the model and a background dataset for integrating out features.

For the first one, we could easily use a *predict* function already implemented in the RS algorithm. As background dataset, we use a sample of the training dataset: using the entire training set could result in a really slow computation of Shapley values.

In order to have an average prediction which does not depend on the sample, we use stratified sampling to maintain the proportion of positive and negative classifications.

Going back to the feature correlation problem, we would like to compute Shapley values for partitions of the coalition and, in particular, we would like to aggregate the discipline levels. The SHAP package includes a *PartitionExplainer* which computes Owen values, but it is still in development and, therefore, not useful for our research.

To address this issue, we decide to use a workaround with the KernelExplainer by providing it with datasets in which discipline levels are grouped. This allows us to compute Shapley values for the aggregated *disciplines* feature while mitigating the high correlation between discipline levels.

| discipline_level_0 | discipline_level_1 | discipline_level_2 | discipline_level_3 |
|---|---|---|---|
| Mathematics and Statistics | Mathematics | Geometry and Topology | Euclidean Geometry |

| disciplines |
|---|
| Mathematics and Statistics/Mathematics/Geometry and Topology/Euclidean Geometry |

Then, we modify the *predict* function so that it ungroups the discipline levels to use all features required by our model.

| disciplines |
|---|
| Mathematics and Statistics/Mathematics/Geometry and Topology/Euclidean Geometry |

| discipline_level_0 | discipline_level_1 | discipline_level_2 | discipline_level_3 |
|---|---|---|---|
| Mathematics and Statistics | Mathematics | Geometry and Topology | Euclidean Geometry |

The SHAP package is most compatible with scikit-learn models and, for the same reason, it requires *One-Hot Encoding* to work with categorical data.

In this case, this type of encoding is not useful, since our model works with categorical features without any type of encoding required.

To overcome this problem, we create a factorized version of both the training set and the dataset of items to explain: this way, we do not have to create a dummy variable for every value of every nonnumeric feature, but we maintain the same number of features and we assign an integer for every value in the non-numeric features. When we encode the training set, we save the "encoding rules" to use the same on other datasets. The following pseudocodes illustrates how the implemented encoding and decoding work.

---

**Algorithm 1** Encode

**Require:** data, encoding_rules=None

    **Check** if discipline levels are grouped or not
    **Assign** the correct feature names

    **if** encoding-rules is not None **then**
        Encode data using encoding_rules
        **return** encoded data

    **else**
        Factorize non-numeric features
        Save encoding_rules
    **end if**
    **return** encoded_data, encoding_rules

---

---

**Algorithm 2** Decode

---

**Require:** encoded_data, encoding_rules

    **Check** if discipline levels are grouped or not
    **Assign** the correct feature names

    Decode encoded_data using encoding_rules
    **return** decoded_data

---

After that, we can define the *new_predict* function.

---

**Algorithm 3** new_predict

---

**Require:** data, model, encoding-rules

    **if** data is encoded **then**
        Decode encoded-data using encoding-rules
    **end if**

    **if** discipline levels are grouped **then**
        Ungroup discipline levels
    **end if**

    predictions ← [ ]
    **for each** row in data **do**
        tmp ← *predict*(row, model)
        predictions.append(tmp)
    **end for**
    **return** predictions

---

We proceed by computing the Shapley values of 100 items from the resource dataset, showing the Feature Importance Plot and evaluating the Explainability with the Keep Absolute (resample) metric.

## 5. RESULTS

First, we show the importance of our model's features:

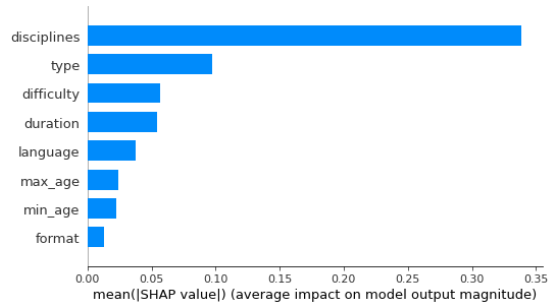

Figure 9: Feature Importance computed using Shapley values.

As can be seen in Figure 9 (and as expected), *disciplines* is by far the most important feature. What makes SHAP's Feature Importance Plot more desirable than classic permutation-based methods is that the importance does not depend on the goodness of the model, but on the magnitude of feature attributions (weighted average of the absolute value of the Shapley values).

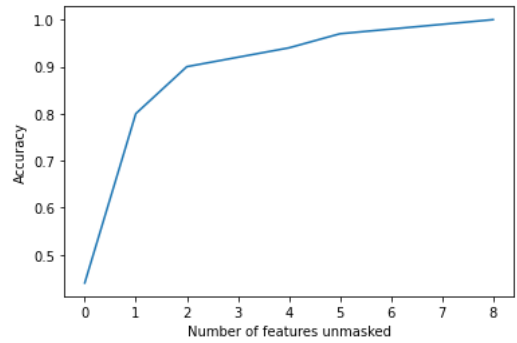

Figure 10: Keep absolute (resample) metric.

The result curve in Figure 10 demonstrates that, with just the first two most important features unmasked, the accuracy reaches 0.90. As the Feature Importance Plot in Figure 9 highlights that *disciplines*, *type*, and *difficulty* are the features with the highest mean of absolute Shapley values, this implies that over 90% of items are accurately recommended or not recommended based on preferences for only these features.

The area under the curve (AUC) for the plotted curve is 0.81, which we denote as ξ.

## 5.1. Total Explainability of Our Recommender System

We calculate the complexity of our model and we obtain an average complexity $\omega_m = 4.06$. And finally, given the value of ξ, we can plot how Total Explainability $Tot(U, \xi)$ evolves as a function of $\omega_b$, considering for $U$ both Eq. 2 and 3.

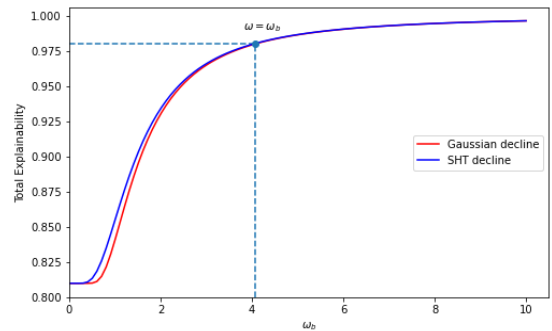

Figure 11: Total Explainability as a function of $\omega_b$.

The pipeline presented made the RS explainable with an explainability value that, depending on $\omega_b$, is in the range $[0.81, 1)$.

## 6. CONCLUSIONS

In this paper, we presented a new quantitative approach to explainable AI (XAI) by evaluating and augmenting the explainability of a Recommender System (RS) in the E-Learning field. We introduced a formal taxonomy to establish a framework for our case study. It is important to note that explainability is not the same as interpretability, which means that even white-box models can benefit from XAI research.

Future research should delve deeper into the applicability of XAI techniques to white-box models and their potential benefits.

There are several potential avenues for future work, including:
−        Conducting studies in Human-Computer Interaction (HCI) to better understand how users manage complexity and, consequently, develop improved understandability functions.
−        Considering user saturation by setting limits on the number of explanations provided, as this could enhance the overall explainability of the system.

# AUTHORS

R Porcedda Bachelor's degree in Physics from Universita di Pisa and, currently, he is pursuing` his Master degree in Data Science from Universita degli studi di Milano-Bicocca. He was a Data` Science Intern in Social Things srl and currently he is CDO (Chief Data Officer) of Liqex Italia Srl. His research interests include XAI (eXplainable Artificial Intelligence), Functional Time Series Analysis and ML&AI applications in Finance.